\documentclass[11pt,a4paper]{elsart}
\usepackage{ifthen,graphics}
\usepackage{color}
\usepackage{cite}
\usepackage[colorlinks]{hyperref}

\definecolor{dgreen}{rgb}{.0,.7,.0}

\newcommand{\Fig}[1]{Fig.~\ref{#1}}
\newcommand{\Ref}[1]{Ref.~\citen{#1}}

\newcommand{\Tab}[1]{Table~\ref{#1}}

\newcommand{\VA}[3]{\ifthenelse{\equal{#2}{#3}}
{\ensuremath{#1\pm#2}}{\ensuremath{#1\,^{+#2}_{-#3}}}}

\newcommand{\Emiss}{\ensuremath{E_\mathrm{miss}}}

\newcommand{\pvec}{\ensuremath{\mathbf{p}}}
\newcommand{\rxy}{\ensuremath{r_{xy}}}

\newcommand{\PK}{\ensuremath{K}}
\newcommand{\PKbar}{\ensuremath{\bar{K}}}

\newcommand{\xc}{\ensuremath{\mathbf{x}_\mathrm{c}}}
\newcommand{\pc}{\ensuremath{\mathbf{p}_\mathrm{c}}}
\newcommand{\lc}{\ensuremath{l_\mathrm{c}}}
\newcommand{\dc}{\ensuremath{d_\mathrm{c}}}

\renewcommand{\vec}[1]{\mbox{\boldmath{$\rm#1$}}}

\newcommand{\pemiss}{\ensuremath{|{\bf p}_{\rm miss}|-E_{\rm miss}}}

\makeatletter
\renewcommand{\section}{\@startsection{section}%
{1}%
{0mm}%
{0.95\baselineskip}
{0.5\baselineskip}
{\normalfont\large\bf\mathversion{bold}}}%
\makeatother
\makeatletter
\renewcommand{\subsection}{\@startsection{subsection}%
{2}%
{0mm}%
{0.95\baselineskip}
{0.5\baselineskip}
{\normalfont\normalsize\bf\mathversion{bold}}}%
\makeatother

\def\ifm#1{\relax\ifmmode#1\else$#1$\fi}  
\def\DAF{DA\char8NE}
\def\x{\ifm{\times}}  
\def\pt#1,#2,{\ifm{#1\x10^{#2}}}

\def\ab{\ifm{\sim}}

\def\to{\ifm{\rightarrow}} 
\def\kl{\ifm{K_L}}   
\def\ks{\ifm{K_S}}
\def\eiii{\ifm{\pi^\pm e^\mp\nu}}  

\def\muiii{\ifm{\pi^\pm \mu^\mp\nu}}   

\def\pio{\ifm{\pi^0\pi^0}} 
\def\po{\ifm{\pi^0}}
\def\pic{\ifm{\pi^+\pi^-}}  

\def\rmk{\rm\kern.5mm }   
\def\f{\ifm{\phi}}

\newcommand{\aff}[2]{Dipartimento di Fisica dell'Universit\`a #1 e Sezione INFN, #2, Italy.}
\newcommand{\affd}[1]{Dipartimento di Fisica dell'Universit\`a e Sezione INFN, #1, Italy.}

\begin{document}
\begin{frontmatter}
\title{\mathversion{bold}Measurement of the Branching Ratio of the
       \kl \to \pic\ decay with the KLOE Detector}
  \pagestyle{plain}
\collab{The KLOE Collaboration}
\author[Na]{F.~Ambrosino},
\author[Frascati]{A.~Antonelli},
\author[Frascati]{M.~Antonelli\corauthref{cor1}},
\author[Roma3]{C.~Bacci},
\author[Karlsruhe]{P.~Beltrame},
\author[Frascati]{G.~Bencivenni},
\author[Frascati]{S.~Bertolucci},
\author[Roma1]{C.~Bini},
\author[Frascati]{C.~Bloise},
\author[Roma1]{V.~Bocci},
\author[Frascati]{F.~Bossi},
\author[Virginia]{D.~Bowring},
\author[Roma3]{P.~Branchini},
\author[Roma1]{R.~Caloi},
\author[Frascati]{P.~Campana},
\author[Frascati]{G.~Capon},
\author[Na]{T.~Capussela},
\author[Roma3]{F.~Ceradini},
\author[Frascati]{S.~Chi},
\author[Na]{G.~Chiefari},
\author[Frascati]{P.~Ciambrone},
\author[Virginia]{S.~Conetti},
\author[Frascati]{E.~De~Lucia},
\author[Roma1]{A.~De~Santis},
\author[Frascati]{P.~De~Simone},
\author[Roma1]{G.~De~Zorzi},
\author[Frascati]{S.~Dell'Agnello},
\author[Karlsruhe]{A.~Denig},
\author[Roma1]{A.~Di~Domenico},
\author[Na]{C.~Di~Donato},
\author[Pisa]{S.~Di~Falco},
\author[Roma3]{B.~Di~Micco},
\author[Na]{A.~Doria},
\author[Frascati]{M.~Dreucci},
\author[Frascati]{G.~Felici},
\author[Frascati]{A.~Ferrari},
\author[Frascati]{M.~L.~Ferrer},
\author[Frascati]{G.~Finocchiaro},
\author[Roma1]{S.~Fiore},
\author[Frascati]{C.~Forti},
\author[Roma1]{P.~Franzini},
\author[Frascati]{C.~Gatti},
\author[Roma1]{P.~Gauzzi},
\author[Frascati]{S.~Giovannella},
\author[Lecce]{E.~Gorini},
\author[Roma3]{E.~Graziani},
\author[Pisa]{M.~Incagli},
\author[Karlsruhe]{W.~Kluge},
\author[Moscow]{V.~Kulikov},
\author[Roma1]{F.~Lacava},
\author[Frascati]{G.~Lanfranchi},
\author[Frascati,StonyBrook]{J.~Lee-Franzini},
\author[Karlsruhe]{D.~Leone},
\author[Frascati]{M.~Martini},
\author[Na]{P.~Massarotti},
\author[Frascati]{W.~Mei},
\author[Na]{S.~Meola},
\author[Frascati]{S.~Miscetti},
\author[Frascati]{M.~Moulson},
\author[Karlsruhe,Frascati]{S.~M\"uller},
\author[Frascati]{F.~Murtas},
\author[Na]{M.~Napolitano},
\author[Roma3]{F.~Nguyen},
\author[Frascati]{M.~Palutan},
\author[Roma1]{E.~Pasqualucci},
\author[Roma3]{A.~Passeri},
\author[Frascati,Energ]{V.~Patera},
\author[Na]{F.~Perfetto},
\author[Roma1]{L.~Pontecorvo},
\author[Lecce]{M.~Primavera},
\author[Frascati]{P.~Santangelo},
\author[Roma2]{E.~Santovetti},
\author[Na]{G.~Saracino},
\author[Frascati]{B.~Sciascia},
\author[Frascati,Energ]{A.~Sciubba},
\author[Pisa]{F.~Scuri},
\author[Frascati]{I.~Sfiligoi},
\author[Frascati]{T.~Spadaro},
\author[Roma1]{M.~Testa\corauthref{cor2}},
\author[Roma3]{L.~Tortora},
\author[Frascati]{P.~Valente},
\author[Karlsruhe]{B.~Valeriani},
\author[Frascati]{G.~Venanzoni},
\author[Roma1]{S.~Veneziano},
\author[Lecce]{A.~Ventura},
\author[Frascati]{R.~Versaci},
\author[Frascati,Beijing]{G.~Xu},
\address[Virginia]{Physics Department, University of Virginia, Charlottesville, VA, USA.}
\address[Frascati]{Laboratori Nazionali di Frascati dell'INFN, Frascati, Italy.}
\address[Karlsruhe]{Institut f\"ur Experimentelle Kernphysik, Universit\"at Karlsruhe, Germany.}
\address[Lecce]{\affd{Lecce}}
\address[Na]{Dipartimento di Scienze Fisiche dell'Universit\`a ``Federico II'' e Sezione INFN, Napoli, Italy}
\address[Energ]{Dipartimento di Energetica dell'Universit\`a ``La Sapienza'', Roma, Italy.}
\address[Roma1]{\aff{``La Sapienza''}{Roma}}
\address[Roma2]{\aff{``Tor Vergata''}{Roma}}
\address[Roma3]{\aff{``Roma Tre''}{Roma}}
\address[Pisa]{\affd{Pisa}}
\address[StonyBrook]{Physics Department, State University of New York at Stony Brook, NY, USA.}
\address[Beijing]{Permanent address: Institute of High Energy Physics, CAS, Beijing, China.}
\address[Moscow]{Permanent address: Institute for Theoretical and Experimental Physics, Moscow, Russia.}

\begin{flushleft}
\corauth[cor1]{cor1}{\small $^1$ Corresponding author: Mario Antonelli
INFN - LNF, Casella postale 13, 00044 Frascati (Roma), 
Italy; tel. +39-06-94032728, e-mail mario.antonelli@lnf.infn.it}
\end{flushleft}
\begin{flushleft}
\corauth[cor2]{cor2}{\small $^2$ Corresponding author: Marianna Testa
Dipartimento di Fisica dell'Universit\`a ``La Sapienza'' e Sezione INFN, P.le Aldo Moro, 2 - 00185 Roma,
Italy; tel. +39-06-49914614, e-mail marianna.testa@roma1.infn.it}
\end{flushleft}

\begin{abstract}
  We present a measurement of the branching ratio of the $CP$ violating
  decay \kl\to\ \pic\ performed by the KLOE experiment at the \f\ factory \DAF.
 We use 328 pb$^{-1}$ of data collected in 2001 and 2002, corresponding to
 $\sim $ 150 million tagged \kl\ mesons.  
 We find
BR(\kl\ \to\ \pic) = $(1.963 \pm 0.012_{\rm stat.} \pm 0.017_{\rm syst.})\times 10^{-3}$.
 This branching ratio measurement is fully inclusive of
final-state radiation. Using the above result, we determine
the  modulus of the amplitude ratio
 $|\eta_{+-}|$ to be $(2.219 \pm 0.013)\times 10^{-3}$ and 
$|\epsilon|$ to be $(2.216 \pm 0.013)\times 10^{-3}$.

\end{abstract}
\end{frontmatter}

\label{sec:br_strategy}
 $CP$ violation was discovered in 1964
 through the observation of the decay \kl \to\\\pic ~\cite{CCFT}.
 The value of BR($\kl\to \pic$) is known today with high accuracy from the 
 results of many experiments, but  a recent measurement by KTeV \cite{KTeV:BrL} is in
 disagreement with the value reported by the PDG \cite{PDG}.
 In the Standard Model, $CP$ violation is naturally accommodated 
 by a phase in the quark mixing matrix\cite{Cabibbo,KM}.
 BR(\kl \to \pic), together with the well known values of BR(\ks \to \pic), $\tau_{\ks}$,
 and $\tau_{\kl}$, determines the modulus of the amplitude ratio
 $|\eta_{+-}|=\break\sqrt{\Gamma(\kl \to \pic)/\Gamma(\ks \to \pic)}$, which is related
 to the $CP$ violation parameters $\epsilon$ and $\epsilon'$ by
$\eta_{+-}=\epsilon+\epsilon'\simeq\epsilon$.
The value of $|\epsilon|$ can thus be obtained and compared with the Standard 
Model calculation. 
  In this letter, we present a measurement of the branching ratio of the
 decay \kl\ \to\ \pic\ performed by the KLOE experiment at the \f\ factory
 \DAF.

The KLOE detector consists of a large, cylindrical drift chamber (DC), surrounded by a 
lead/scintillating-fiber electromagnetic calo\-ri\-meter (EMC).
 A superconducting coil around the calorimeter 
provides a 0.52 T field. The drift chamber \cite{KLOE:DC} is 4~m in diameter and 3.3~m long.
The momentum resolution is $\sigma_{p_{\perp}}/p_{\perp}\approx 0.4\%$. 
Two-track vertices are reconstructed with a spatial resolution of $\sim$ 3~mm. 
The calorimeter \cite{KLOE:EmC} is divided into a barrel and two endcaps.
 It covers 98\% of the solid angle.
 Cells close in time and space are grouped into calorimeter clusters.
 The energy and time resolutions for photons of energy $E$ are
 $\sigma_E/E = 5.7\%/\sqrt{E\ {\rm(GeV)}}$
 and $\sigma_t = 57\ {\rm ps}/\sqrt{E\ {\rm(GeV)}}\oplus100\ {\rm ps}$, respectively.
For this analysis, events triggered \cite{KLOE:trig} only by calorimeter signals are used. Two energy deposits above threshold ($E>50$ MeV for the barrel and  $E>150$ MeV for the endcaps) are required.

For the present measurement, we use a subset of the data collected by 
KLOE during the years 2001 and 2002 that satisfies basic quality 
criteria \cite{KLOE:brlnote}. The data used corresponds to an integrated 
luminosity of \ab328 pb$^{-1}$.
Each run used in the analysis is simulated with the KLOE Monte Carlo  
program, \textsc{GEANFI} \cite{KLOE:offline}, using values of relevant machine 
parameters such as $\surd s$ and $\vec p_\f$ as determined from data.
Because of the small beam-crossing angle, \f\ mesons are produced at  \DAF\  with a momentum 
of about of $12$ MeV/$c$, toward the ring's center. 
Machine background obtained from data is superimposed on Monte Carlo events
on a run-by-run basis. The number of events simulated for each run in the 
data set is equivalent to that expected on the basis of the run luminosity. 
For this analysis, we use a Monte Carlo sample consisting of $\f\to\ks\kl$ events
in which the \ks\ and \kl\ decay in accordance with their natural 
branching ratios. The effects of initial- and final-state radiation are included in the
 simulation. The treatment of final-state radiation in \ks\ and \kl\ decays 
 follows the method discussed in Ref.~\citen{KLOE:gattip}.
 In particular, the \kl\to\pic($\gamma$) event generator includes
 the amplitudes for inner bremsstrahlung and direct emission.
Good runs, and the corresponding Monte Carlo events, are 
organized into 14 periods.
The branching ratio analysis is performed independently for each of these
periods. The average over the 14 periods gives the final result.

 At a \f\ factory, neutral kaons are produced through  $\f\to\ks\kl$ 
 decays. A pure sample of \kl's can be selected by identification of
 $\ks\to\pic$  decays (tagging).
 The tagging efficiency depends slightly on the fate of the \kl:
 it is different for events in which the \kl\ decays to each channel,
 interacts in the calorimeter, or escapes the detector.
 KLOE has already used  the tag method to measure the dominant \kl\
 branching ratios and the \kl\ lifetime\cite{KLOE:brl}.
 The tagging criteria described in \Ref{KLOE:brl} where chosen to
 minimize the difference in the tagging efficiency among the various decay modes. 
 In the present analysis, in order to increase the statistics, we employ a tagging
 algorithm with an efficiency of about a factor of seven larger than that used in \Ref{KLOE:brl}.

 A precise measurement of BR$(\kl \to \pic)$ is obtained here
 by measuring the ratio  $R = {\rm BR}(\kl \to \pic)/{\rm BR}(\kl \to \muiii)$ since the values of
 of the tagging efficiency for  \kl\to\pic\ and \kl\to$\pi\mu \nu$  are similar.
 The value of BR$(\kl \to \pic)$ is finally obtained using the value of 
 BR$(\kl \to \muiii)$ from \Ref{KLOE:brl}.

For \kl\ tagging, we select $\ks\to\pic$ decays by requiring two tracks of opposite
 curvature from a vertex in a cylindrical FV with $\rxy < 10$~cm and $|z| < 20$~cm, 
 centered on the collision region as determined for each run using
 Bhabha events.
We also require that the two tracks give $|m(\pic)-m_{K^0}|<5$ MeV/$c^2$ and that 
$||\vec p_++\vec p_-|-p_{K_S}|<$ 10 MeV/$c$, with $p_{K_S}$ calculated from
the kinematics of the $\f\to\ks\kl$ decay.
The \kl\ momentum, $p_{\kl}$, 
is obtained from the \ks\ direction and $\vec p_\f$.
The \kl\ line of flight ({\em tagging line}) is then constructed from the \kl\ momentum, 
$\vec p_{\kl} = \vec p_\f-\vec p_{\ks}$, and the position of the production 
vertex, $\vec x_\f$.

 The overall value of the tagging efficiency is about 66\%, and that  
 of the ratio of the tagging efficiencies for events in which 
 \kl\to\pic\ and \kl\to$\pi\mu \nu$, averaged over all periods, is $1.012 \pm  0.001$
  as obtained from Monte Carlo.

 To reconstruct the \kl\ decay vertex,
 all relevant tracks in the chamber,
 after removal of those from the \ks\ 
decay and their descendants, are extrapolated to their points of closest
 approach to the  tagging line. 
For each track candidate, we evaluate the point of closest approach to the 
tagging line, \xc, and the distance of closest approach, \dc. The
momentum \pc\ of the track at \xc\ and the extrapolation length, \lc, are also
computed. Tracks satisfying $\dc< a \rxy + b$, with $a=0.03$ and $b=3$ cm, and 
$-20$ cm $<\lc<25$ cm are accepted as \kl\ decay products, \rxy\ being  the
 distance of \xc\ from the origin in the transverse plane.
 For each sign of charge we consider as associated to the \kl\ decay the
 track with the smallest value of \dc. We attempt to find a vertex using these
 two tracks. Candidate \kl\ vertices must be within the  fiducial volume
 defined by $30<\rxy<150$ cm and $|z|<120$ cm. 
 
The tracking efficiency has been evaluated by Monte Carlo 
with corrections obtained from data control samples\cite{KLOE:brlnote}.
The corrections range between 0.99 and 1.03 depending on the \kl-decay channel 
and on the run period. We determine the conditional 
tracking efficiency as the ratio of the number of
events in which there are two \kl-decay tracks of opposite sign
and the number of events in which there is a single \kl-decay track.
We use the same method for both data and Monte Carlo,
and correct the efficiency obtained from the simulation
 with the data-Monte Carlo ratio of the conditional 
tracking efficiencies from control samples.
The corrections are evaluated as functions of the track momentum using 
\kl \to \pic\po\ and \kl\to $\pi e \nu$ control samples. 
To select \kl \to \pic\po\ events, we require at least 
one \kl\ decay track and two photons from  $\pi^0$ decay.
To reject background, we apply  cuts on the two-photon invariant mass
and on the photon times of flight.
The momenta of decay tracks are calculated with a resolution of about 10 MeV/$c$
using the two photons from the $\pi^0$ decay and the momentum of the
other track. In order to obtain the corrections for higher momentum tracks
(above $\sim$150 MeV/$c$),  \kl\to $\pi e\nu$ decays are used.  
To select \kl\to $\pi e \nu$ decays, we identify electrons by time of flight
 with a purity of about 95\%. The background is mostly due to $K_{\mu 3}$ decays.
The momentum of the second \kl\ track is evaluated from the missing 
energy with a resolution of about 30 MeV/$c$. 
The value of the  tracking efficiency is about 61\% for \pic\ decays and 51\% for \muiii\
 decays.
Once two tracks have been found, the vertex efficiency  is about 97\%. 
We find values for data and Monte Carlo to be in agreement within 0.1\%.

 The number of $\kl\to\muiii (\gamma)$ events is obtained by  fitting  the
 distribution of $\Delta_{\pi\mu}=\pemiss$ with a linear combination of Monte Carlo
 distributions as in \Ref{KLOE:brl}.  $\Delta_{\pi\mu}$ 
 is obtained from the smaller absolute value of the two possible values of \pemiss,
 where $\vec p_{\rm miss}$ and 
 \Emiss\ are the missing momentum and missing energy at the \kl\ vertex, evaluated
 by assuming the two tracks to be a $\pi\mu$ or $\mu\pi$ pair.

\begin{figure}[ht]
  \begin{center}
    \resizebox{0.49\textwidth}{!}{\includegraphics{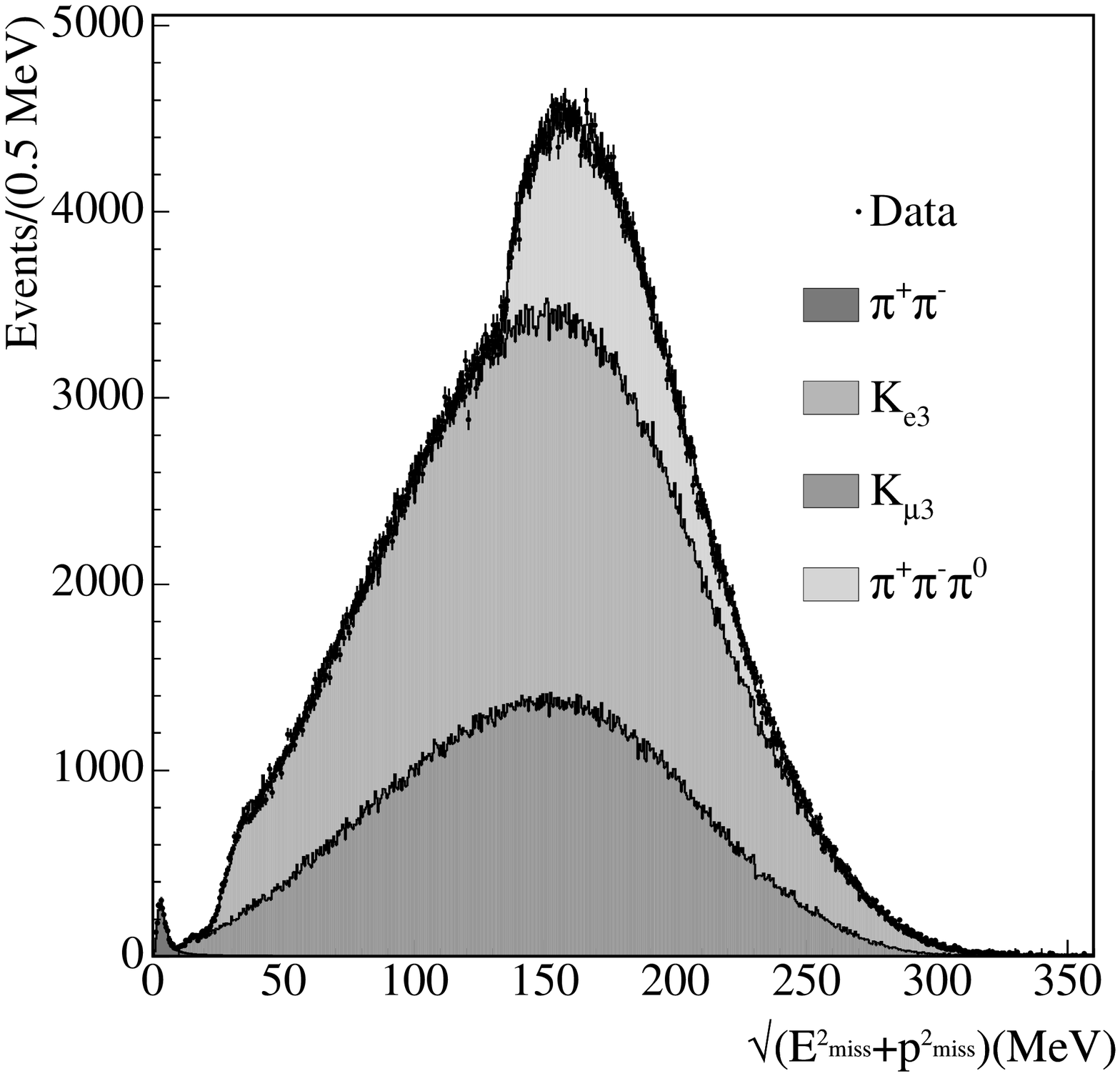}}
    \resizebox{0.49\textwidth}{!}{\includegraphics{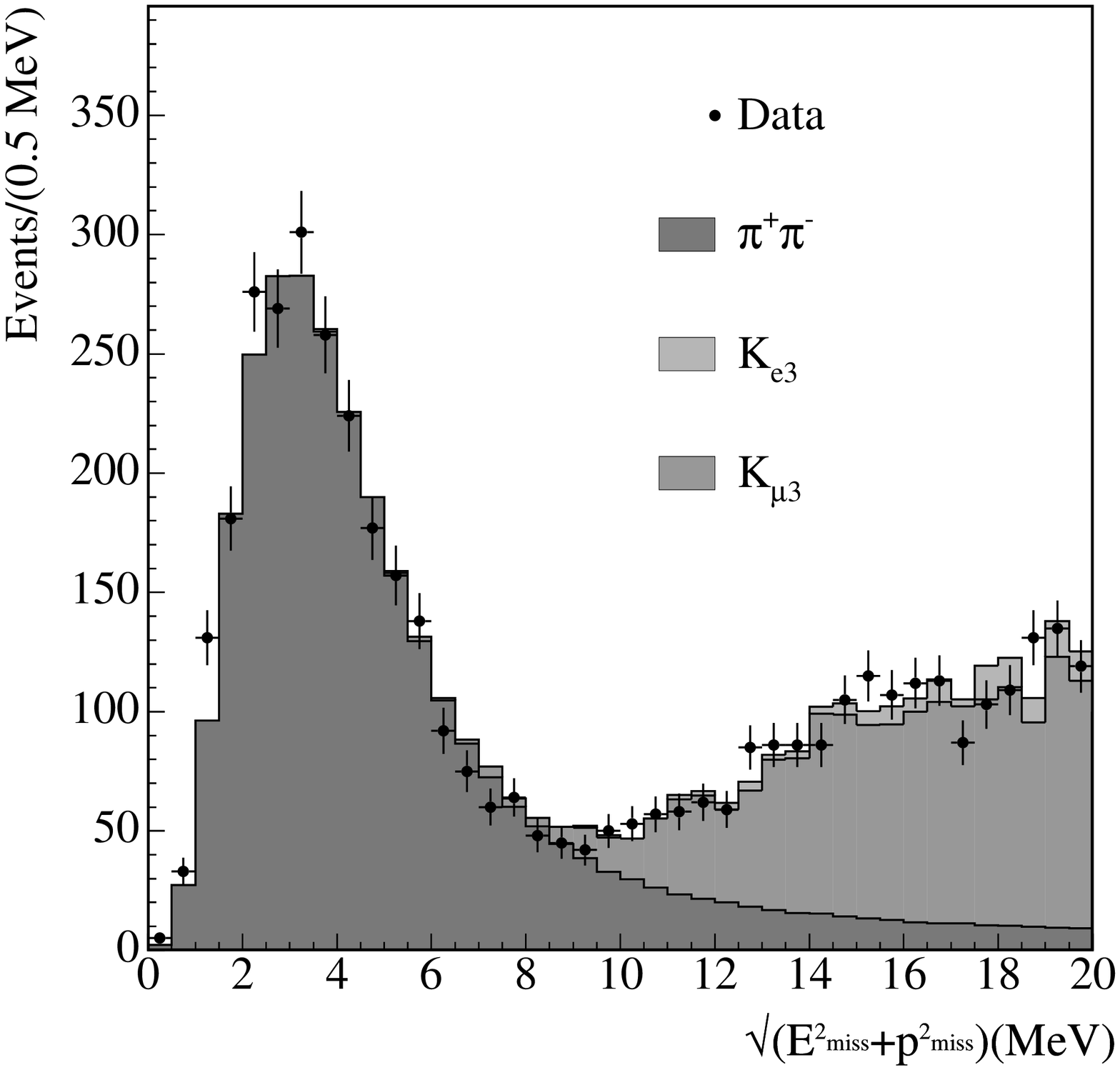}}
  \end{center}
  \caption{Distribution of $\sqrt{E^2_{\rm miss}+|\pvec_{\rm miss}|^2}$ for 
a single run period, with fit to Monte Carlo distributions for different decay channels:
 full fit range (left panel); signal region (right panel).}
  \label{fig:spectra_ip10_pp}
\end{figure}

The best variable for selecting $K_L \to \pic (\gamma)$ decays is 
$\sqrt{E^2_{\rm miss}+|\pvec_{\rm miss}|^2}$ where 
 $E_{\rm miss}$ in this case is the missing energy in the hypothesis
 of the $\kl\ \to \pic$ decay.
The result of the fit to the $\sqrt{E^2_{\rm miss}+|\pvec_{\rm miss}|^2}$
 distribution with a linear combination of Monte Carlo
 distributions  is shown in
 \Fig{fig:spectra_ip10_pp}, left for a single run period.
The signal region is shown expanded in \Fig{fig:spectra_ip10_pp}, right.
 The sum over the 14 periods of the numbers of signal and normalization events obtained 
 from the fits,
 and the average values over the 14 periods for the efficiency ratios
 are given in \Tab{tab:sel_summary}.
\begin{table*}[htbp!]
  \begin{center}
    \begin{tabular}{|c|c|}
\hline
number of \pic\ events   & 45267 $\pm$ 255 \\
number of $\pi \mu \nu$ events   & 5243723 $\pm$ 2805 \\
$\epsilon_{{\rm tagging}}(\pic)/\epsilon_{{\rm tagging}}(\pi \mu \nu)$ & 1.0119 $\pm$ 0.0008 \\
$\epsilon_{{\rm tracking}}(\pic)/\epsilon_{{\rm tracking}}(\pi \mu \nu)$ & 1.1726 $\pm$ 0.0011 \\

\hline
\end{tabular}
\caption{Analysis summary: number of signal and normalization events obtained from the fits,
 and the average values of the efficiency ratios.}
\label{tab:sel_summary}
\end{center}
\end{table*}


 The systematic errors due to the limited knowledge of the 
 corrections, data-Monte Carlo discrepancies, and instabilities in the signal
counting have been studied as described in the following.

The systematic error on the determination of the ratio of the tagging efficiencies
is evaluated from the the stability of the results obtained 
with different tagging criteria. 
Since the variations in the tagging efficiency are mostly due to the
 dependence of the calorimeter trigger on the \kl\ decay channel, the
 analysis has been repeated with the additional requirement that the event trigger
 be due to the \ks\ decay products alone ({\it self-triggering} event). 
Another contribution to the systematic uncertainty on the tagging efficiency is 
due to the dependence of the reconstruction efficiency of the
pions from $\ks \to \pic$ on the presence of other tracks in the drift chamber.
This effect is reduced for events in which the pion directions are almost orthogonal
 to the \ks\ line of flight. The analysis has been repeated with the requirement that
 the difference between the \ks\  pion momenta be less than 
 60 MeV. The largest fractional difference in the results obtained using different 
tagging criteria is $32 \times 10^{-4}$. This value is taken as the systematic uncertainty
 on the tagging efficiency.

The uncertainty on the tracking efficiency is dominated by
the control sample statistics and by the variation of the results
 observed by using different criteria to identify tracks from  \kl\ decays.
The systematic error due to the possible bias introduced
in the selection of the control sample has been studied by varying 
the values of the cuts made on \dc\ and \lc\ when associating tracks to \kl\ vertices.
These changes result in a variation of the tracking efficiency
by about $\pm$20\%. The corresponding  fractional change in $R$ is
$37\times 10^{-4}$. This value is taken as the systematic uncertainty on the
 tracking efficiency.

 The systematic error due to the reliability of the Monte Carlo distributions 
 in the variables  \pemiss\ and $\sqrt{E^2_{\rm miss}+|\pvec_{\rm miss}|^2}$ has been evaluated
 by studying the sensitivity of $R$ to the momentum resolution.
 The fits to the $\sqrt{E^2_{\rm miss}+|\pvec_{\rm miss}|^2}$ and the \pemiss\
 distributions have been repeated using Monte Carlo samples in which the
 momentum resolution has been varied according to the uncertainty
 determined using high purity control samples of the dominant \kl\ decays. 
 A fractional systematic uncertainty of $55\times 10^{-4}$  is obtained. 

\begin{figure}[h!]
  \begin{center}
    \resizebox{0.7\textwidth}{!}{\includegraphics{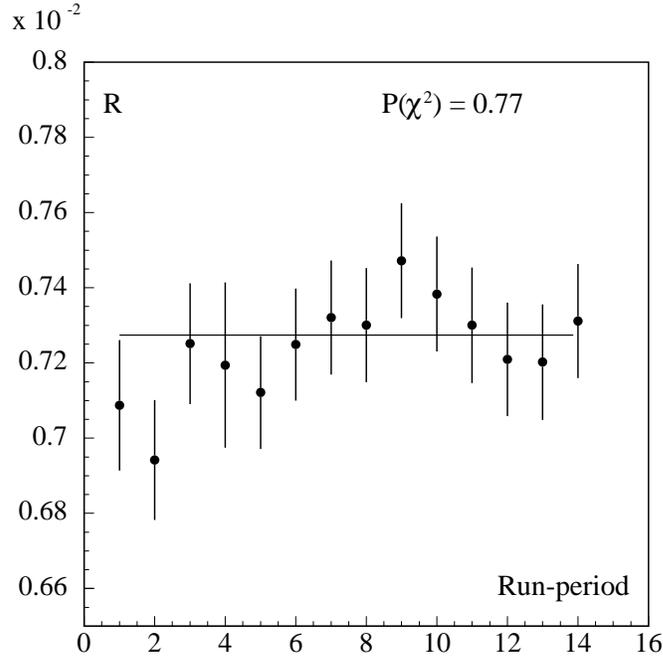}}
  \end{center}
  \caption{Values of R for 14 run-periods. The result of a fit to a constant value is
 also shown.}
  \label{fig:stab}
\end{figure}

 The final value for BR$(K_L \to \pic)$ has been calculated as the average over
 the 14 periods of data taking. Good stability of the results is observed
 after application of the period-dependent tracking efficiency corrections.
 The $\chi^2$ value of the fit corresponds to a confidence level of 77\%;
 see \Fig{fig:stab}.
Including systematic uncertainties, we obtain:
 \begin{eqnarray*}
\frac{{\rm BR}(\kl \to \pic)}{{\rm BR}(\kl \to \pi \mu \nu)}  
=(0.7275  \pm  0.0042_{\rm stat}\pm 0.0054_{\rm syst} ) \times 10^{-2} 
\end{eqnarray*}
where the statistical error includes the contribution from Monte Carlo statistics
and the statistical uncertainty on the tracking efficiency correction.  
 Using the KLOE result BR($K_L \to \pi \mu \nu$)=0.2698$\pm$0.0015 \cite{KLOE:brl} 
we find:
 \begin{eqnarray*}
 {\rm BR} (K_L \to \pic) &=& (1.963 \pm 0.012_{\rm stat.} \pm 0.017_{\rm syst.})\times 10^{-3}
\end{eqnarray*}
 This result is fully inclusive with respect 
  to final-state radiation including both the inner bremsstrahlung and the 
($CP$-conserving) direct emission components.
 The result is in good agreement with the measurement from KTeV \cite{KTeV:BrL},
\footnote{The contribution of direct emission is claimed to be negligible 
  in the KTeV result.}
$(1.975\pm 0.012)\times 10^{-3}$, and in 
 strong disagreement with the value reported by the PDG, 
 $ (2.090\pm 0.025)\times 10^{-3}$ \cite{PDG}.

 The error on the measurement of  BR($K_L \to \pi \mu \nu$) of \Ref{KLOE:brl}
  is correlated with the errors  on
 BR($K_L\to\pi e \nu$),  BR($K_L \to \pio \po$),
 BR($K_L \to  \pic \po$), and the \kl\ lifetime.
 This is because, in the treatment of \Ref{KLOE:brl}, the branching ratios and the
 lifetime are determined from 
 lifetime-dependent absolute branching ratio measurements using the constraint
 $\sum {\rm BR} = 1-0.0036$, where 0.0036 represents the small contribution due to 
 the sum of the branching ratios for  $K_L \to \pic$,  $K_L \to \pio$,
 and  $K_L \to \gamma \gamma$ listed in the PDG compilation\cite{PDG}.
 As a consequence, the error on the value for ${\rm BR} (K_L \to \pic)$
 reported here is correlated with the errors on the branching ratio 
 and \kl\ lifetime values reported in \Ref{KLOE:brl}. 
 Additional sources of correlation arise from common systematic
 uncertainties in the branching ratio measurements.
 The complete correlation matrix is:
\begin{displaymath}
\left(
\begin{array}{cccccc}
 1 & -0.25 & -0.56 & -0.07 &-0.12 & 0.25 \\
   &     1 & -0.43 & -0.20 &0.51 & 0.33 \\
   &       & 1     & -0.39 &-0.24 & -0.21 \\
   &       &       & 1     &-0.09 & -0.39 \\
   &       &       &       & 1& 0.11\\     
   &       &       &       & & 1\\     
\end{array}
\right)
\end{displaymath}
 where the columns and the rows refer, in order, to the
 BR's for decays to $\eiii$, $\muiii$, $\pio\po$, $\pic\po$ and
 $\pic$, and $\tau_{\kl}$.  
 Because ${\rm BR} (K_L \to \pic)$ is small, the KLOE values for the
 dominant BR's are essentially unchanged when the new KLOE value for
 ${\rm BR} (K_L \to \pic)$ is used in constraining the sum of the
 \kl\ BR's to unity.

\begin{figure}
\begin{center}
 \resizebox{0.7\textwidth}{!}{\includegraphics{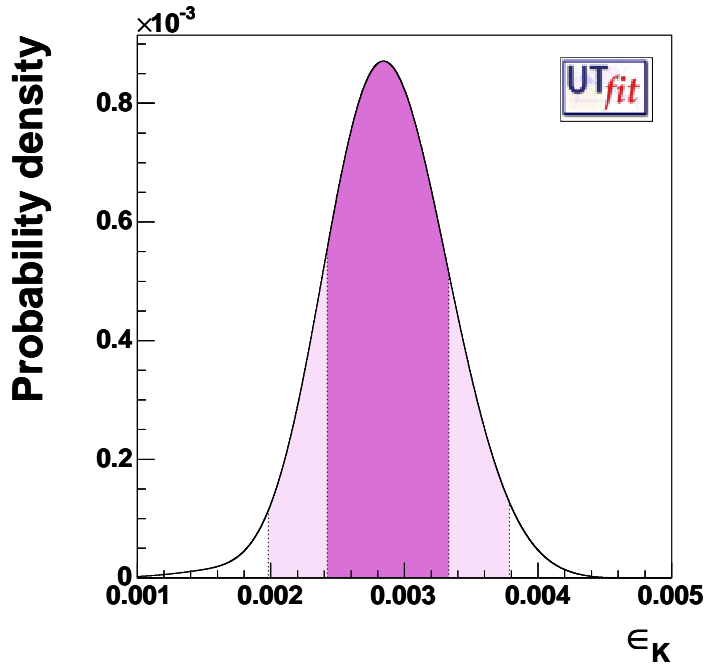}}
  \end{center}
\caption{  Standard Model prediction of $|\epsilon|$ from  measurements 
 of $|V_{ub}|/|V_{cb}|$, and $\Delta M(B_d)$ and from the limit on
 $\Delta M(B_s)$ from \Ref{UTfitweb}.
}
\label{fig:epskco}
\end{figure}

The measurement of BR($\kl\to \pic$) can be used to determine $|\eta_{+-}|$
and $|\epsilon|$, correcting for the small contribution of $\epsilon'$.
Using the measurements of BR($\ks\to \pic$) and $\tau_{\kl}$ from KLOE
\cite{KLOE:Rs,KLOE:brl}  and the value of  $\tau_{\ks}$ from the 
 PDG\cite{PDG}, and subtracting the contribution of the $CP$-conserving direct-photon
 emission process $\kl\to \pic \gamma$
\cite{Dire} from the inclusive measurement of BR($\kl\to \pic$), we obtain
 $|\eta_{+-}| = (2.219 \pm 0.013)\times 10^{-3}$.
Finally, using the world average measurement of 
$\rm{Re}(\epsilon'/\epsilon)=(1.67 \pm 0.26)\times 10^{-3}$ 
 and assuming $\arg\epsilon'=\arg\epsilon$, we obtain 
$|\epsilon| = (2.216 \pm 0.013)\times 10^{-3}$.
 This result is 
 in disagreement with the value $|\epsilon| = (2.284 \pm 0.014)\times 10^{-3}$ 
 reported in the PDG compilation~\cite{PDG}.
 The value of $|\epsilon|$ can be 
 be predicted from the measurement of other observables 
 and compared with measurements.
 For this purpose, we use the prediction 
 $|\epsilon| = (2.875 \pm 0.455)\times 10^{-3}$ obtained by the UTfit collaboration
 ~\cite{UTfitweb,UTfit}, 
 where to test the mechanism of the $CP$ violation in the Standard Model,
 the value of $|\epsilon|$ has been 
 computed from  measurements of the $CP$-conserving observables 
 $\Delta M(B_d)$, $\Delta M(B_s)$,
 $V_{ub}$, and  $V_{cb}$. 
 The probability density function of  the prediction of $|\epsilon|$ 
 is shown in  \Fig{fig:epskco}.
 No significant deviation from the Standard Model prediction is
 observed. Notice that, due to the large uncertainties on the computation
 of the hadronic matrix element corresponding to the \PK-\PKbar\ mixing,
 the theoretical error is much larger than the experimental error.
 
 In summary, a new, precise measurement of the  \kl\ \to\ \pic\ branching ratio has been
 performed. The result:
 \begin{eqnarray*}
 {\rm BR} (K_L \to \pic) &=& (1.963 \pm 0.012_{\rm stat.} \pm 0.017_{\rm syst.})\times 10^{-3}
\end{eqnarray*}
is in  good agreement  with that recently obtained by KTeV \cite{KTeV:BrL},
 and in disagreement with the value reported by the PDG\cite{PDG}.
The impact of this measurement on the determination of the apex of the unitarity triangle
 has been studied and compared with determinations from measurements 
 of other observables.

\section*{Acknowledgments}
We thank warmly M. Pierini for fruitful discussions on the interpretation of the UTfit
 results for the prediction of $|\epsilon|$.
We thank the DA$\Phi$NE team for their efforts in maintaining low background running 
conditions and their collaboration during all data-taking. We want to thank our technical staff: 
G.F. Fortugno for his dedicated work to ensure an efficient operation of the KLOE Computing Center; 
M. Anelli for his continuous support to the gas system and the safety of the detector; 
A. Balla, M. Gatta, G. Corradi and G. Papalino for the maintenance of the electronics;
M. Santoni, G. Paoluzzi and R. Rosellini for the general support to the detector; 
C. Piscitelli for his help during major maintenance periods.
This work was supported in part by DOE grant DE-FG-02-97ER41027; 
by EURODAPHNE, contract FMRX-CT98-0169; 
by the German Federal Ministry of Education and Research (BMBF) contract 06-KA-957; 
by Graduiertenkolleg `H.E. Phys. and Part. Astrophys.' of Deutsche Forschungsgemeinschaft,
Contract No. GK 742; by INTAS, contracts 96-624, 99-37; 
by TARI, contract HPRI-CT-1999-00088.

\bibliographystyle{elsart-num}
\bibliography{kl-pp}

\end{document}